\documentclass[12pt,a4paper]{JHEP}
\usepackage{amsmath,amssymb}

\title{Perturbative and non-perturbative aspects of pure
  $\mathcal{N}=1$ super Yang-Mills theory from wrapped branes}

\author{W.~M\"uck \\
  Dipartimento di Scienze Fisiche, Universit\`a di Napoli
  ``Federico II'', \\ 
  and INFN, Sezione di Napoli, Via Cintia, 80125 Napoli, Italy\\
  E-mail: \email{mueck@na.infn.it}}

\abstract{The Maldacena-Nu\~nez solution is generalized to include a
  number of integration constants, one of which controls the
  resolution of the singularity of the wrapped D5-brane
  background. Some features of the dual pure $\mathcal{N}=1$
  super Yang-Mills (SYM) theory are calculated, amongst which the gluino
  condensate, the beta function of the gauge coupling and a brane
  probe potential, which is related to the Veneziano-Yankielowicz
  effective potential. Each integration constant has a precise meaning
  in the dual SYM theory, e.g., the amount of non-perturbative SYM
  physics captured by the gravity configuration is described by the
  singularity resolution parameter.}


\preprint{DSF-3-2003\\hep-th/0301171}

\providecommand{\e}{}
\renewcommand{\e}[1]{\mathrm{e}^{{#1}}}
\newcommand{\tth}{\tilde{\theta}}
\newcommand{\tp}{\tilde{\phi}}
\newcommand{\hth}{\hat{\theta}}
\newcommand{\hp}{\hat{\phi}}
\newcommand{\da}{\dot{a}}
\newcommand{\hr}{\hat{\rho}}
\newcommand{\gYM}{g_{\text{YM}}}
\newcommand{\tYM}{\theta_{\text{YM}}}
\newcommand{\tvac}{\theta_0}
\newcommand{\glue}{\langle \lambda^2\rangle} 
\newcommand{\VVY}{V_{\mathrm{VY}}}
\newcommand{\Weff}{W_{\mathrm{eff}}}
\newcommand{\rL}{\rho_\Lambda}
\newcommand{\trace}{\,\mathrm{Tr}\,}
\begin{document}

\section{Introduction and Summary}
\label{intro}
The possibility of studying super Yang-Mills (SYM) theories using
their gravity duals has been a surprising manifestation of 't~Hooft's
old idea that gauge theories have a string theoretical microscopic
origin \cite{'tHooft:1974jz}. After the success of the AdS/CFT
correspondence \cite{Maldacena:1998re,Gubser:1998bc,Witten:1998qj}
(see also the recent lecture notes \cite{D'Hoker:2002aw,Skenderis:2002wp})
in describing the large $N$ limit of (super) conformal SYM theories
(\emph{e.g.}, $\mathcal{N}=4$ SYM theory) by asymptotically AdS super gravity
(SUGRA) backgrounds, a systematic formulation of a more general
``gauge/gravity'' duality describing also non-conformal SYM theories (or gauge
theories with fewer supersymmetries) is still an outstanding
problem. A huge amount of work has recently been devoted to the study
of specific cases of this duality, and much relevant information of the
SYM theories has been extracted from their SUGRA duals. One way to
reduce supersymmetry is to consider SUGRA backgrounds
generated by branes wrapping supersymmetric cycles
\cite{Bershadsky:1996qy,Maldacena:2000mw}. 
Thus, the identification of such a SUGRA dual of
pure $\mathcal{N}=1$ SYM theory by Maldacena and Nu\~nez (MN)
\cite{Maldacena:2000yy} was a major achievement and has spurred a lot
of activity.\footnote{Another dual of pure $\mathcal{N}=1$ SYM theory
is the Klebanov-Strassler solution \cite{Klebanov:2000hb} describing
fractional 3-branes at the apex of a deformed conifold.} For a list of 
references to other known cases of the gauge/gravity correspondence,
we refer the reader to \cite{DiVecchia:2002ks}.

The MN solution was originally found as a BPS magnetic monopole by
Chamseddine and Volkov \cite{Chamseddine:1997nm,Chamseddine:1998mc}, 
and since MN's work it has been the subject of a number of articles. A
qualitative analysis of its implications for
$\mathcal{N}=1$ SYM theory has been performed by Loewy and Sonnenschein
\cite{Loewy:2001pq}. The importance of the gaugino condensate for
de-singularizing the SUGRA solution was discussed by Apreda, Bigazzi,
Cotrone, Petrini and Zaffaroni \cite{Apreda:2001qb}. 
Building on this observation, Di~Vecchia, Lerda
and Merlatti \cite{DiVecchia:2002ks,Merlatti:2002uz,DiVecchia:2002gw} 
used a D5-brane embedded in the MN background and wrapping
a certain two-cycle in order extract the running of the gauge coupling and the
$\tYM$ angle. In particular, they found agreement to leading order
with the Novikov-Shifman-Vainshtain-Zakharov (NSVZ) beta function
\cite{Novikov:1983uc,Novikov:1986rd} and evidence for
fractional instantons. However, their specific way of wrapping the
D5-branes led to a number of problems, most notably, their full beta
function contains terms, which are logarithmic in the coupling and
cannot be interpreted in field theory. Although the two-loop
coefficient of the beta function can be adjusted by changing the
radial/energy relation \cite{Olesen:2002nh}, the problem of the
logarithmic terms remained. Its resolution involves a
suitable change of renormalization scheme, which could be translated
to a change of the two-cycle around which the D5-brane is wrapped.   
In fact, Bertolini and Merlatti
\cite{Bertolini:2002yr} solved this problem by wrapping the D5-brane
on a different cycle, which had already been indicated in the paper by
MN.

Other aspects of the MN solution, which have been studied in the
literature, include the resolution of the conifold singularity using
black holes \cite{Buchel:2001qi}, a different approach to the
radial/energy relation \cite{Wang:2002ka}, non-supersymmetric
deformations \cite{Aharony:2002vp,Evans:2002mc,Apreda:2003gc}, 
its non-commutative extension \cite{Mateos:2002rx} and its Penrose
(pp-wave) limit \cite{Gomis:2002km,Correa:2002sp,Gimon:2002nr}.

In this paper, we shall again consider the MN solution and add a
number of new ingredients to the discussion. Thus, we hope not only to
contribute to a better understanding of this specific case of the
gauge/gravity duality, but also to provide a guideline for
analyzing other cases. 
 
Let us now summarize our paper and interpret the main results. 
One of the main ideas of the gauge/gravity duality is that there
exists a dictionary relating the SUGRA fields to certain SYM
operators. Taking the simplest approach, we generalize the MN
solution by allowing for three integration constants, which have a
precise physical meaning in the dual SYM theory. These are the
following. First, we introduce a constant angle, $\psi_0$, 
by performing in the MN 
solution a global rotation of the frame of the twisted
three-sphere. In the SYM theory, $\psi_0$ is identified with the phase
of the gluino condensate. Moreover, it also 
determines the vacuum angle, $\tvac$, which is the value of $\tYM$ at the 
vacuum. For each $\tvac$, there are $N$ inequivalent values of
$\psi_0$ giving rise to the $N$ physically inequivalent vacua. Second,
the dilaton constant $\Phi_0$ relates the value of the dynamically
generated SYM scale $\Lambda$ to the string parameters by the relation
$\e{2\Phi_0} = 2(2\pi)^4 (\Lambda/L)^3 /(N^3 g_s^2)$, where $L$ is the
SUGRA scale given by $L^{-2}= N g_s\alpha'$. This result stems from a
direct calculation of the gluino condensate.
Finally, we include an integration constant $c$, which can
be found in the solution by Gubser, Tseytlin and Volkov
\cite{Gubser:2001eg}, and which controls the resolution of the
singularity. In fact, for $c>0$, the bulk geometries possess a bad 
naked singularity, whereas the case $c=0$ corresponds to the regular
MN solution. The most natural interpretation of this constant is that
it measures the amount of non-perturbative SYM physics captured by the
dual SUGRA geometry. For $c=\infty$, the SUGRA solution contains only
perturbative effects, whereas the MN solution describes also the
non-perturbative physics. This interpretation is supported by the
running of the gauge coupling, the value of the gluino condensate and
the behaviour of the probe brane potential. The (generalized) MN
solution is reviewed in Sec.~\ref{review}.

The MN solution is of the form $\mathbb{R}^{1,3}\times M_6$, where $M_6$
is a (non-compact) Calabi-Yau manifold. Its geometry encodes a variety
of SYM quantities
\cite{Gukov:1999ya,Taylor:1999ii,Mayr:2000hh,Dijkgraaf:2002dh}. Of
these, we calculate in Sec.~\ref{gluino_pot} the gluino condensate and
the effective superpotenial. The gluino condensate is a constant,
which, for the regular MN solution, we identify with $\Lambda^3$, where
$\Lambda$ is the dynamically generated scale of the SYM theory
\cite{Davies:1999uw}. 

The remaining sections deal with the application of the brane probe
technique, which is laid out in Sec.~\ref{probe}.\footnote{The
  embedded object carrying the D5-brane action 
  will be called a brane \emph{probe}, although the usual zero-force condition
  cannot be imposed. Instead, we shall try to interpret the probe
  potential as an effective potential of the dual SYM theory.}
We obtain the gauge coupling and $\tYM$ angle and, by considering the
terms of the 
probe action that are independent of the gauge fields, the probe potential.
Our analysis generalizes the calculation of \cite{Bertolini:2002yr}, in that
we do not fix the angular coordinate of the embedding. Thus, we obtain
the correct (and in general non-vanishing) $\tYM$. In
contrast, the wrapping of \cite{DiVecchia:2002ks} yields a
$\tYM$ differing from our result by a factor $1/2$ (with a somewhat 
difficult interpretation of the chiral symmetry), while the
result of \cite{Bertolini:2002yr} corresponds to the special case $\tYM=0$.

Sec.~\ref{betafunc} focuses on the analysis of the gauge coupling
obtained by the brane probe, but the breaking of the chiral
symmetry by perturbative and non-perturbative effects ($U(1)\to
\mathbb{Z}_{2N}\to \mathbb{Z}_{2}$) shall also be discussed. The main
result will be the calculation of the beta function. In contrast to
\cite{Bertolini:2002yr} we shall average the probe gauge coupling over all
inequivalent vacua in order to confront its running 
with a perturbative field theory analysis. This will not only remove a
spurious energy dependence from $\tYM$, but we will also be able to
\emph{exactly} re-write the beta function in terms of gauge theory
quantities. It will turn out that the (singular) solution with
$c=\infty$ correctly yields the complete perturbative running in terms
the NSVZ beta function, and non-perturbative
effects appear in terms of a ($c$-dependent) value of the gluino
condensate, which might be re-written in terms of fractional instanton 
contributions in the far UV. Thus, our beta function predicts
new terms, which have not been obtained in the field theory. In
fact, the pole of the NSVZ beta function seems to disappear in the complete
theory, which is dual to the (regular) MN solution. For the
singular solutions, the location of the pole of the beta function
coincides with the minimum of the effective superpotential from the
Calabi-Yau geometry, which is thus interpreted as a vacuum-averaged
result. 

While the gauge coupling is relevant for the microscopic (UV) degrees
of freedom, the probe potential provides a measure of the effective
degrees of freedom around the vacuum. We shall analyze it in detail in
Sec.~\ref{probepot}. For the singular solutions, the
probe brane will fall into the singularity, and the state of lowest
energy appears to be chirally invariant. Fortunately, the perturbative
analysis of the coupling breaks down before the brane reaches the
singularity. In contrast, for the regular case, the minimum of the
probe potential is not chirally invariant, and we obtain a good
description of the behaviour of the composite operator $\lambda^2$,
where $\lambda$ is the gluino field, around the vacuum. In the same
region the probe potential will turn out to be closely related to the
Veneziano-Yankielowicz effective potential \cite{Veneziano:1982ah}.

\section{Review of the MN solution}
\label{review}
We consider the SUGRA solution corresponding to a system of $N$
D5-branes. One way of finding it is by using $d=7$
gauged SUGRA, which is obtained as a consistent truncation of the
$d=10$ SUGRA by compactification on an $S^3$
\cite{Cvetic:2000dm,Chamseddine:1999uy}. This is a natural setting to
incorporate the twist condition necessary to retain some supersymmetry
\cite{Bershadsky:1996qy,Maldacena:2000mw}. 
The metric in the string frame is \cite{Maldacena:2000yy,DiVecchia:2002ks}
\begin{equation}
\label{metric10}
  ds_{10}^2 = \e{\Phi} \left[ dx_{1,3}^2 + \frac{\e{2h}}{L^2}
  (d \tth^2 +\sin^2 \tth \, d\tp^2) +\frac1{L^2} d\rho^2
  +\frac1{L^2} \sum\limits_{a=1}^3 (\sigma^a-L A^a)^2
  \right]~.
\end{equation}
Here, $\tth \in [0,\pi)$ and $\tp \in [0,2\pi)$ parameterize a
two-sphere, $S^2$, which is part of the gauged SUGRA solution. The 
compactification three-sphere, $S^3$, is parameterized by the
left-invariant one-forms $\sigma^a$ ($a=1,2,3$) and is twisted by the
$SU(2)$ gauge field $A=\tau^a A^a$, where $\tau^a$ denote
the Pauli matrices. For completeness, we give here the
expressions for the $\sigma^a$,
\begin{align}
\label{sigma1}
  \sigma^1 &= \frac12 [ \cos(\psi-\psi_0) \, d\theta 
  + \sin(\psi-\psi_0) \sin\theta \, d\phi ]~,\\
\label{sigma2}
  \sigma^2 &= \frac12 [ -\sin(\psi-\psi_0) \, d\theta 
  + \cos(\psi-\psi_0) \sin\theta \, d\phi ]~,\\
\label{sigma3}
  \sigma^3 &= \frac12 [ d\psi + \cos\theta \, d\phi ]~,
\end{align}
which satisfy
\begin{equation}
\label{dsigma}
  d\sigma^a = - \varepsilon^{abc} \sigma^b \wedge \sigma^c~.
\end{equation}
The angles are defined in the intervals
$\phi\in[0,2\pi)$, $\theta\in [0,\pi)$ and $\psi\in
[0,4\pi)$. In addition to the solution in the literature, 
we have included the constant angle $\psi_0$, 
which arises from a global $U(1)$ gauge transformation. 
More precisely, one could consider the
transformed gauge field $A'= g^{-1} A g +i g^{-1} dg$, where $g\in
SU(2)$. Then, the part of the metric that belongs to the twisted 3-sphere
can be written in the form (apart from the warp factor)
\begin{equation}
  \sum\limits_{a=1}^3 (\sigma^a -L {A'}^a)^2 
  = \frac12 \trace ( \sigma -L A' )^2 
  = \frac12 \trace (g \sigma g^{-1} -iL dg \,g^{-1} -L A)^2~.
\end{equation}
Hence, a \emph{global} gauge transformation corresponds to a pure rotation
of the frame on $S^3$, while \emph{local} transformations will, in
addition to the rotation, contribute to the twisting. Our frame is
obtained from the MN frame by the transformation
$g=\exp (-i\psi_0\tau^3/2)$. 

The dilaton $\Phi$ and the prefactor $\e{2h}$ in the metric are functions of
the radial variable $\rho$ and are given by
\begin{align}
\label{Phi}
  \e{2\Phi} &= \e{2\Phi_0} f(c) \frac{\sinh(2\rho+c)}{2\e{h}}~,\\ 
\label{h}
  \e{2h} &= \rho \coth(2\rho+c) -\frac14 [a(\rho)^2+1]~,\\
\label{a} 
  a(\rho) &= \frac{2\rho}{\sinh(2\rho+c)}~.
\end{align}
In eqn.~\eqref{Phi}, $f(c)$ is part of the overall constant, but we
choose not to absorb it into $\Phi_0$. The reason for this is that we want to
consider $c$ and $\Phi_0$ as \emph{independent} integration constants
related to distict features of the dual gauge theory. We impose
that for the MN solution ($c=0$), $f(0)=1$. Moreover, if the solution
with $c=\infty$ and finite $\Phi_0$ is to make sense, we also need
$f(c)\sim\e{-c}$ for $c\to\infty$. $f(c)$ shall be determined in Sec.\
\ref{betafunc}. 

The $SU(2)$ gauge fields, $A^a$, are given by
\begin{equation}
\label{A}
  A^1 = \frac{a(\rho)}{2L}  d\tth~,\quad
  A^2 = \frac{a(\rho)}{2L}  \sin\tth \, d\tp~,\quad
  A^3 = \frac1{2L} \cos \tth \, d\tp~,
\end{equation}
with the field strengths $F^a = dA^a + L \epsilon^{abc} A^b
\wedge A^c$, 
\begin{equation}
\label{F}
  F^1 = \frac{\da(\rho)}{2L} d\rho \wedge d \tth~,\quad
  F^2 = \frac{\da(\rho)}{2L} d\rho \wedge \sin\tth \,d\tp~,\quad
  F^3 = \frac1{2L} \left[a(\rho)^2-1\right] d\tth \wedge \sin\tth \, d\tp~.
\end{equation}
The dot denotes a derivative with respect to $\rho$.

Furthermore, the solution contains a 2-form potential
\begin{equation}
\label{C2}
\begin{split}
  C^{(2)} &= \frac1{4L^2} \left[ \psi \left(\sin\theta \, d\theta \wedge
  d\phi) + \sin\tth \, d\tth \wedge d\tp \right) + \cos\theta \cos\tth
  \, d\phi \wedge d\tp \right] \\
  &\quad - \frac{a(\rho)}{2L^2} \left( d\tth \wedge \sigma^1 +
  \sin\tth \, d\tp \wedge \sigma^2 \right)~, 
\end{split}
\end{equation}
whose 3-form field strength $F^{(3)} = dC^{(2)}$ is 
\begin{equation}
\label{F3}
  F^{(3)} = \frac2{L^2} (\sigma^1-L A^1) \wedge
  (\sigma^2-L A^2) \wedge (\sigma^3-L A^3) 
  - \frac1{L} \sum\limits_{a=1}^3 F^a \wedge  
  (\sigma^a-L A^a)~.
\end{equation}

The metric \eqref{metric10} is real for $\rho\ge\hr$, where $\hr$ is
defined by $\e{2h(\hr)}=0$. It is not difficult to show that $\hr$ is
implicitly determined by the transcedental equation
\begin{equation}
\label{hatrhodef}
  2\hr\left[ \coth(2\hr+c) +1\right]  = 1~,
\end{equation}
which has a unique solution $0\le \hr \le 1/4$. The limiting cases are 
$\hr=0$ for $c=0$ and $\hr=1/4$ for $c=\infty$.

The constant $L$ is related to the number $N$ of wrapped
D5-branes by the usual charge quantization condition
\begin{equation}
\label{charge_quant}
  \frac1{2\kappa_{10}^2} \int_{S^3} F^{(3)} = N \tau_5~.
\end{equation}
Using $\kappa_{10}=8\pi^{7/2} g_s {\alpha'}^2$ and $\tau_5=(2\pi)^{-5}
g_s^{-1} {\alpha'}^{-3}$ one obtains \cite{DiVecchia:2002ks}
\begin{equation}
\label{lambda}
  L^{-2} = N g_s \alpha'~.
\end{equation}

In addition to the fields listed so far, there is a non-zero 6-form
potential, $C^{(6)}$, defined by $dC^{(6)}=\star F^{(3)}$, where the Hodge
dual is taken with respect to the string frame metric \eqref{metric10}.
Using eqns.~\eqref{metric10}, \eqref{A}, \eqref{F} and \eqref{F3}, it
is straightforward to obtain
\begin{equation}
\label{dC6}
\begin{split}
  dC^{(6)} &= -\frac1{L^2} v^{(4)} \wedge \left\{ 2 \e{2\Phi}
  \left[\e{2h} +\frac1{16} \e{-2h} (1-a^4) \right] d\rho\wedge
  d\tth \wedge \sin\tth\, d\tp \phantom{\frac12} \right. \\
  &\quad \left. + \frac14 d\left[ 
  \e{2\Phi} \da \left( d\tth \wedge \sigma^2 - \sin \tth\, d\tp \wedge
  \sigma^1 \right) 
  - (a^2-1) \e{2\Phi-2h} d\rho \wedge (\sigma^3-L A^3) \right] \right\}~,
\end{split}
\end{equation}   
where we have abbreviated $v^{(4)}=dx^1\wedge dx^2\wedge dx^3\wedge
dx^4$. 
Thus, the 6-form potential is 
\begin{equation}
\label{C6}
\begin{split}
  C^{(6)} &= -\frac1{L^2} v^{(4)} \wedge \left\{ 
  \Psi(\rho)\, d\tth \wedge \sin\tth\, d\tp \right. \\
  &\quad \left. + \frac14 \left[ 
  \e{2\Phi} \da \left( d\tth \wedge \sigma^2 - \sin \tth\, d\tp \wedge
  \sigma^1 \right) 
  - (a^2-1) \e{2\Phi-2h} d\rho \wedge (\sigma^3 -L A^3) \right]
  \right\}~, 
\end{split}
\end{equation}   
where the function $\Psi(\rho)$ satisfies
\begin{equation}
\label{dotPsi}
  \dot{\Psi}(\rho) =  2 \e{2\Phi}
    \left[\e{2h} +\frac1{16} \e{-2h} (1-a^4) \right]~.
\end{equation}
We were not able to integrate this equation, except for the case
$c=\infty$, where $\Psi=\e{2\Phi}(\rho-1/2)+\Psi_0$, with $\Psi_0$
being an integration constant. We shall comment further on the
function $\Psi$ in Sec.~\ref{probepot}.
\section{Gluino condensate and effective superpotential}
\label{gluino_pot}
Let us start our analysis by considering the geometry of the
``internal'' manifold. The bulk solution is---apart from the warp
factor---of the form $\mathbb{R}^{1,3}\times M_6$, where $M_6$ is a 
K\"ahler manifold and geometrically encodes various aspects of the
dual gauge theory. It encodes, first, the effective superpotential 
\begin{equation}
\label{superpot_def}
  \Weff \sim \int_{M_6} F^{(3)} \wedge \Omega~,
\end{equation} 
where $\Omega$ is the holomorphic 3-form of the complex manifold
$M_6$. Since $M_6$ is not compact, $\Weff$ explicitly depends on a
cut-off. The holomorphic 3-form $\Omega$ is given by
\cite{Ivanov:2000fg,Papadopoulos:2000gj}\footnote{We
have chosen the phase so that $W_{\mathrm{eff}}$ is real.}
\begin{equation}
\label{form3}
  \Omega = \frac{\e{2\Phi}}{L^3} 
  \mathcal{E}^1 \wedge \mathcal{E}^2 \wedge \mathcal{E}^3~,
\end{equation}
where the complex 1-forms $\mathcal{E}^i$ are defined by
\begin{align}
\label{E1}
  \mathcal{E}^1 &= (\sigma^3-L A^3) -i d\rho~,\\
  \mathcal{E}^2 &= (\sigma^1-L A^1)+iX \e{h} \sin\tth d\tp -iP
  (\sigma^2-L A^2)~,\\
  \mathcal{E}^3 &= \e{h} d\tth +iX ( \sigma^2-L A^2) +iP \e{h}
  \sin \tth d\tp~,
\end{align}
and 
\begin{equation}
\label{PX}
  P = \frac{\sinh(4\rho+2c)-4\rho}{2\sinh^2(2\rho+c)}~,\qquad 
  X = (1-P^2)^{1/2} = \frac{2 \e{h}}{\sinh(2\rho+c)}~.
\end{equation}
A straightforward calculation yields 
\begin{equation}
\label{superpot}
  \Weff(\rho_0) \sim \frac{16 \pi^3}{L^5} \e{2\Phi_0} f(c)
  \int\limits_{\hr}^{\rho_0} d\rho \left[ 2\rho \coth(2\rho+c) -1
  \right]~.
\end{equation}

The effective superpotential can be recast in terms of a pre-potential after
introducing a canonical basis of homology 3-cycles of $M_6$
\cite{Taylor:1999ii,Mayr:2000hh},
\begin{equation}
  \Weff \sim \int_A F^{(3)} \int_B \Omega - \int_A \Omega \int_B F^{(3)}~.
\end{equation}  
In our case the compact 3-cycle is $A=S^3$, and the non-compact 3-cycle $B$ has
a complicated form. Since we have already found $\Weff$, we shall
consider only the compact $S^3$.
The integral of $F^{(3)}$ over $S^3$
is proportional to the number of D-branes, $N$, see eqn.\
\eqref{charge_quant}, while the integral of $\Omega$ over $S^3$ encodes the gluino condensate,
\begin{equation}
\label{gluino_cond}
  2\pi i |\glue_c| = \tau_5 \int_{S^3} \Omega =
  i \tau_5 \frac{2\pi^2}{L^3} \e{2\Phi} X 
  = i \tau_5 \frac{2\pi^2}{L^3} \e{2\Phi_0} f(c)~. 
\end{equation}
The constant $\tau_5$ is needed for dimensional reasons.
Following our interpretation of the integration constants, we re-write
eqn.\ \eqref{gluino_cond} as 
\begin{equation}
\label{gluino_c}
  |\glue_c| = |\glue_0| f(c)= \Lambda^3 f(c)~,
\end{equation}
where we have used the
convention $f(0)=1$ and the fact that the regular solution is the true
dual of $\mathcal{N}=1$ SYM theory, \emph{i.e.} $\glue_0=\Lambda^3$,
where $\Lambda$ is the dynamically generated mass scale.
Thus, we identify the precise role of $\Phi_0$ relating $\Lambda$ to
the SUGRA parameters by the relation 
\begin{equation}
\label{Phi0_interpret}
  \e{2\Phi_0} = \pi^{-1} \tau_5^{-1} \Lambda^3 L^3 = 
  \frac{2(2\pi)^4}{N^3 g_s^2} \left(\frac{\Lambda}{L}\right)^3~.
\end{equation}
Obviously, for $c=\infty$ we have $|\glue_\infty| = 0$,
in agreement with the fact that a purely perturbative calculation
fails to exhibit the gluino condensate.

\section{Brane probe analysis}
\label{probe}
Another way of obtaining information about the dual field theory is
by using the probe technique. 
Let us consider a D5-brane embedded in the background
\eqref{metric10}. Its action is given by 
\begin{equation}
\label{D5action}
  S = - \tau_5 \int d^6 \xi\, \e{-\Phi} \sqrt{-\det(G+2\pi\alpha' F)} 
  + \tau_5 \int \left( \sum\limits_n C^{(n)} \wedge
  \e{2\pi\alpha' F} \right)_{\text{6-form}}~.
\end{equation}
We consider a D5-brane wrapping a two-sphere parameterized by two
angles $\hth$ and $\hp$. Expanding the Born-Infeld part of
the action \eqref{D5action} and demanding that the non-abelian gauge
fields $F$ live only in the 4d part of the D5-branes, one finds
\begin{equation}
\label{D5action_exp}
  S = - \int d^4x \left[ V + \frac1{4\gYM^2} F^A_{\mu\nu} F_A^{\mu\nu} 
  -\frac{\tYM+2\pi n}{32\pi^2}  F^A_{\mu\nu} (\star F_A)^{\mu\nu} \right]~,
\end{equation}
where the raising of the indices and the dual of the gauge fields are
taken using the 4d Minkowski metric, and we have used the convention
$\mathrm{tr}(T^A T^B)=\frac12 \delta^{AB}$ for the colour trace over the
non-abelian generators. The potential $V$ is given by 
\begin{equation}
\label{Vdef}
  V = \tau_5 \int d\hth d\hp\, \left( \e{-\Phi} \sqrt{-G} -
    C^{(6)}_{1234\hth\hp} \right)~.
\end{equation}  
For the gauge coupling, $\gYM$, and
the theta angle, $\tYM$, one obtains \cite{DiVecchia:2002ks} 
\begin{align}
\label{gYMdef}
  \frac1{\gYM^2} &= 2 \pi^2 {\alpha'}^2 \tau_5 \int d\hth d\hp\,
  \e{-3\Phi} \sqrt{-\det G}~,\\
\label{tYMdef}
  \tYM &= (2 \pi)^4 {\alpha'}^2 \tau_5 \int d\hth d\hp\,
  C^{(2)}_{\hth\hp} \mod 2\pi~.
\end{align}
In eqns.\ \eqref{D5action_exp} and \eqref{tYMdef} we have used the fact
that the physics of Yang-Mills theory is periodic in the theta angle with
period $2\pi$, and we adopt the convention $\tYM\in [0,2\pi)$. 
The metric, the 2-form and the 6-form are induced from the respective
bulk fields. 

In order to proceed we have to specify how the world volume
coordinates of the D5-brane are related to the bulk coordinates of the
MN solution. The flat 4d part is obvious, but the wrapped $S^2$ needs
some care. In order to use the coordinates $\rho$ and $\psi$ as
parameters, we have to ensure that both of them are trivially fibred
over the world volume \cite{Maldacena:2000yy}. This is done by
imposing the four embedding conditions\footnote{The alternative
  $\theta = -\tth$ is physically equivalent.} 
\begin{equation}
\label{D5cond}
  \theta = \tth=\hth~,\quad \phi =\tp=\hp~,\quad 
  \psi=\text{const}~,\quad \rho=\text{const}~.
\end{equation}
Thus, we have $d\rho=0$ and $\sigma^3-L A^3=0$
on the world volume.
Notice that the first two conditions differ from the ones used in
\cite{DiVecchia:2002ks}, where $\theta$ and $\phi$ are kept constant.
Hence, the induced metric on the world volume of the D5-branes becomes
\begin{equation}
\label{metric6}
  ds_6^2 = \e{\Phi} \left\{ dx_{1,3}^2 + \frac1{4L^2}
  \left[ 4\e{2h} +a(\rho)^2 +1 -2a(\rho)\cos(\psi-\psi_0)\right] 
  \left( d\hth^2 +\sin^2\hth \, d\hp^2 \right) \right\}~,
\end{equation}
so that
\begin{equation}
\label{sqrtG}
  \sqrt{-G} = \e{3\Phi} \sin\hth \frac1{L^2}
  \left[ \rho \coth(2\rho+c) - \frac12 a(\rho) \cos(\psi-\psi_0)\right]~.
\end{equation}
Moreover, the induced 2- and 6-forms are 
\begin{align}
\label{C2ind}
  C^{(2)} &= \frac1{2L^2} \sin\hth \, d\hth \wedge d\hp \left[
  \psi- a(\rho) \sin(\psi-\psi_0)\right]~,\\
\label{C6ind}
   C^{(6)}_{1234\hth\hp} &= -\frac1{L^2} \sin\hth \left[
   \Psi(\rho) +\frac14 \e{2\Phi}\da \cos(\psi-\psi_0) \right]~,
\end{align}
respectively.

Inserting these equations into the general expressions for $V$, $\gYM$
and $\tYM$, we find 
\begin{equation}
\label{V}
  V= \frac{4\pi \tau_5}{L^2} \left[ 
  \e{2\Phi} \rho \coth(2\rho+c) + \Psi(\rho) - \frac14 \e{2\Phi}(
  2a -\da) \cos(\psi-\psi_0) \right]~,
\end{equation}
as well as 
\begin{align}
\label{gYM}
  \frac1{\gYM^2} &= \frac{N}{4\pi^2}   
 \left[ \rho \coth(2\rho+c) -\frac12 a(\rho) \cos(\psi-\psi_0)\right]~,\\
\label{tYM}
  \tYM &= N \left[ \psi - a(\rho) \sin(\psi-\psi_0) \right] \mod 2\pi~.
\end{align}

\section{Beta function}
\label{betafunc}
In this section we shall analyze and interpret the gauge coupling and
$\tYM$ angle measured by a probe D5-brane. Our main interest lies in the
calculation of the perturbative beta function, but other aspects, such
as the breaking of chiral symmetry from classically $U(1)$ to
$\mathbb{Z}_{2N}$ in the perturbative regime and to $\mathbb{Z}_2$ by
non-perturbative effects, will also become transparent.

Let us start by discussing the chiral symmetry. We argue that the
transformation $\delta\psi=-2\epsilon$, where $\epsilon \in [0,2\pi)$,
corresponds to a chiral transformation of the dual SYM theory. The
perturbative physics is correctly captured by the solution with
$c=\infty$, in which case eqns.\ \eqref{gYM} and \eqref{tYM} simplify
to  
\begin{equation}
\label{YMinf}
  \frac1{\gYM^2} = \frac{N}{4\pi^2} \rho \quad \text{and} \quad 
  \tYM = N\psi \mod 2\pi~.
\end{equation}
Moreover, in this solution the non-abelian gauge field becomes abelian by virtue of $a(\rho)=0$, which removes all terms with
$\sin(\psi-\psi_0)$ and $\cos(\psi-\psi_0)$ from the metric and the
form fields. Hence, the metric and the field strengths $dC^{(2)}$
and $dC^{(6)}$ are symmetric under $\delta\psi=-2\epsilon$ for all
$\epsilon \in [0,2\pi)$. These transformations form the classical
chiral symmetry group $U(1)$. 
However, $\tYM$ is determined by $C^{(2)}$, which is not
invariant under a general chiral transformation. In fact, eqn.\
\eqref{YMinf} is invariant only for $\epsilon=\pi (n-1)/N$ for
$n=1\ldots 2N$, corresponding to the group $\mathbb{Z}_{2N}$ of the
non-anomalous chiral symmetry transformations. 
In contrast, for every solution with $c<\infty$, where a certain
amount of non-perturbative effects are included, terms with
$\sin(\psi-\psi_0)$ and $\cos(\psi-\psi_0)$ appear showing that
the symmetry of the bulk solutions is given by 
$\epsilon=\pi$ only, which represents
the generator of the unbroken $\mathbb{Z}_2$ chiral symmetry of the
quantum theory. We clearly see that the breaking $\mathbb{Z}_{2N} \to
\mathbb{Z}_2$ is a non-perturbative effect.

Let us turn now to the beta function. Considering the case $c=\infty$,
which is given by eqn.\ \eqref{YMinf}, one is tempted to identify
$\e{2\rho} \sim \mu^3$, where the exponent of $\mu$ is chosen such
that the correct coefficient of the one-loop beta function is
reproduced \cite{Apreda:2001qb}. We shall present an 
alternative argument, which is applicable for any solution with
$c<\infty$. To begin, let us combine $\gYM$
and $\tYM$ to the complexified gauge coupling,
\begin{equation}
\label{tau}
  \tau = \frac{\tYM}{2\pi} +i \frac{4\pi}{\gYM^2} 
  =i \frac{N}{2\pi} \left[ 2\rho \coth(2\rho+c) -i\psi -a(\rho)
  \e{-i(\psi-\psi_0)} \right]~.
\end{equation}
In order to identify the energy scale $\mu$, we follow
\cite{Apreda:2001qb,DiVecchia:2002ks} 
and interpret the function $a(\rho)$ as the
gluino condensate measured in units of $\mu$,\footnote{A different
approach to finding a radial/energy relation can be found in
\cite{Wang:2002ka}.} 
\begin{equation}
\label{mudef}
  a(\rho) = \frac{|\glue_c|}{\mu^3}~,
\end{equation}
where $|\glue_c|=\Lambda^3 f(c)$ by virtue of eqn.\ \eqref{gluino_c}.
We shall in the following be able to determine the function $f(c)$. 
The identification \eqref{mudef} is ambiguous in the case
$c=\infty$, but the result of the following arguments will have a
well-defined limit for $c\to\infty$, because any non-zero $|\glue_c|$
drops out when calculating the beta function.  

The complex coupling \eqref{tau} or, alternatively, eqns.\ \eqref{gYM}
and \eqref{tYM}, might be interpreted as exact, non-perturbative
expressions. More precisely, starting at very large $\rho$ and
a certain initial $\psi$, the RG flow proceeds along the direction of
steepest descent of the probe potential $V$ (see Sec.\ \ref{probepot})
towards the 
minimum at $\psi=\psi_0$, $\rho=\hr$. In the regular case, $c=0$, the
gauge coupling, $\gYM$, diverges at $\rho=0$, signalling the
disappearance of the UV degrees of freedom at the vacuum. 
This exact RG flow was analyzed in \cite{Bertolini:2002yr} for the
special case $\psi=\psi_0$.

The exact RG flow ``knows'' about the position of the vacuum at
$\psi=\psi_0$ in the sense that the value of $\tYM$ flows towards
$\tvac=N\psi_0 \mod 2\pi$. In contrast, perturbative calculations
in field theory are typically ignorant about the vacuum state. For
example, an argument as to why a direct perturbative calculation of 
the gaugino condensate yields
zero is that the perturbative analysis averages over all vacua
\cite{Amati:1985uz,Amati:1988ft}.
We wish to confront our results with a perturbative field theory
calculation and, therefore, we must average over all inequivalent vacua, which
can be equivalently expressed as the following change 
of renormalization scheme,\footnote{It is irrelevant whether we
  average over the $N$ discrete values of $\psi_0$ for a given
  $\tvac$, or whether we take the average over the whole interval.}
\begin{equation}
\label{tauredef}
  \tau\to \tau +i\frac{N}{2\pi} 
  \frac{\glue_c}{\mu^3 \e{i\psi}}~,
\end{equation}
where we have written $\glue_c= |\glue_c| \e{i\psi_0}$, leading to the
coupling 
\begin{align}
\label{gYM_new}
  \frac1{\gYM^2} &= \frac{N}{4\pi^2} \rho \coth(2\rho+c)~,\\
\intertext{and the $\theta$-angle}
\label{tYM_new}
  \tYM &= N \psi \mod 2\pi.
\end{align}
It is interesting to observe that, in this renormalization scheme, the
coupling \eqref{gYM_new} does not diverge for $\rho=c=0$, in contrast
to the exact coupling \eqref{gYM}. 

From eqns.\ \eqref{gYM_new} and \eqref{mudef} it is straightforward to
obtain the beta function of the gauge coupling, 
\begin{equation}
\label{beta}
 \beta = \mu \frac{d\gYM}{d\mu} 
  = -\frac{3N}{16\pi^2} \gYM^3 
  \left[ 1-a(\rho)^2\frac{N\gYM^2}{8\pi^2} \right]
  \left[1-\frac{N\gYM^2}{8\pi^2} \right]^{-1}~.
\end{equation}
For $c\to\infty$, where non-perturbative effects are absent, this
coincides with the complete perturbative beta function of NSVZ
\cite{Novikov:1983uc,Novikov:1986rd}. Notice also that in obtaining 
\eqref{beta} no leading order approximation has been made in order to
rewrite $\rho$ in terms of $\gYM$, and thus \eqref{beta} holds at all
energies. Non-perturbative effects are described by $a(\rho)$, which,
in a large $\rho$ expansion, takes the form of instanton
corrections. In fact, $a(\rho) \approx 16\pi^2/(N\gYM^2)
\exp [-8\pi^2/(N\gYM^2)] \e{-c}$. We see that the constant $c$ determines
how much instanton physics is present in the running. 
Notice, however, that these non-perturbative effects have been
obtained via a ``perturbative'' calculation, in that we are
considering the vacuum-averaged coupling.

By using eqn.\ \eqref{mudef}, eqn.\ \eqref{beta} is recast in terms
of gauge theory quantities only,
\begin{equation}
\label{beta_new}
 \beta  = -\frac{3N}{16\pi^2} \gYM^3 
  \left[1- \frac{|\glue_c|^2}{\mu^6} \frac{N\gYM^2}{8\pi^2} \right]
  \left[1-\frac{N\gYM^2}{8\pi^2} \right]^{-1}~.
\end{equation}
In this way we are also able to determine $f(c)$. In fact, consider
the cases $c>0$, where the beta function diverges for $\gYM^2=8\pi^2/N$
or, equivalently, for $\rho=\rL$, where $\rL$ is determined by 
\begin{equation}
\label{rhoLambda}
  2 \rL \coth( 2\rL +c) =1~.
\end{equation}
Clearly, we have $\rL>\hr$. For the limiting cases, $c=\infty$ and
$c=0$, we find $\rL=1/2$ and $\rL=0$, respectively. (The regular case
$c=0$, in which $\rL=\hr=0$, will be considered below.) Thus, the
perturbative analysis breaks down before the brane reaches the
singularity. Defining $\Lambda$ as the scale
where the beta function diverges, we
obtain from eqn.\ \eqref{mudef} that 
\begin{equation}
\label{f_def}
  f(c) = a(\rL) = \sqrt{1-4\rL^2}~,
\end{equation}
where $\rL$ is determined implicitly by eqn.\ \eqref{rhoLambda}. 
It is straightforward to check that $f(0)=1$ and, for $c\to\infty$,
$f(c)\approx 2\e{-(c+1)}$, as needed. 

In the regular case the beta function is not
singular at $\rho=\rL=0$, because also the numerator is zero. More
precisely, the limit is 
\begin{equation}
\label{beta_limit}
  \beta\to -\frac{3N}{8\pi^2} \gYM^3 
  = -3 \left(\frac{8\pi^2}{N}\right)^{1/2}~.
\end{equation}
Hence, the running changes from NSVZ type for very large $\mu$
to pure one-loop at $\mu=\Lambda$. Notice, however, the factor $2$
with respect to the UV one-loop coefficient.

An interesting observation is that $\rL$ coincides with the location
of the minimum of the effective superpotential, $\Weff$, cf.\
eqn.~\eqref{superpot}. Thus, we are lead to interpret $\Weff$ as a
``perturbative'' expression that averages over all inequivalent vacua.

\section{Probe potential}
\label{probepot}
The vacuum of $\mathcal{N}=1$ SYM is usually described by the
Veneziano-Yankielowicz effective potential, $S \log S$, where $S$ is a
chiral superfield containing the composite operator $\lambda^2$, where
$\lambda$ is the gluino field, as its lowest component. After
integrating out the auxiliary fields, this becomes 
\cite{Veneziano:1982ah},
\begin{equation}
\label{V_VY}
  \VVY \sim |\phi|^{4/3} \left[ \ln^2 \frac{|\phi|}{\Lambda^3} 
  + \left( \alpha -\frac{\tvac+2\pi n}{N} \right)^2 \right]~.
\end{equation}
In the notation of \cite{Veneziano:1982ah},
$\mu^3=\Lambda^3\e{i\tvac/N}$, where $\tvac$ 
is the vacuum angle. The integer $n$ stems from the fact that we may
choose a branch of the logarithm $\ln[(\phi/\mu^3)^N]$
\cite{Kovner:1997im}.
The scalar field $\phi=|\phi|\e{i\alpha}=\lambda^2$ 
describes the effective degrees of freedom at low
energies. Since $\alpha$ has period $2\pi$, there are $N$ inequivalent
values of $n$. Hence, the minimum of $\VVY$ at
$\phi=\Lambda^3\e{i(\tvac+2\pi n)/N}$ describes the $N$ inequivalent
vacua of $\mathcal{N}=1$ SYM. The chirally symmetric minimum at
$|\phi|=0$ is unphysical, since the second derivative of $\VVY$
diverges. 

We would like to find a SUGRA derived quantity that can be compared to
$\VVY$. The effective superpotential \eqref{superpot} is not a good
candidate, because it does not contain an angular variable that could
play the role of $\alpha$. We shall argue in the following that the
probe brane potential $V$ can be compared to a potential derived from
$\VVY$ in the vicinity of the vacuum. 
Consider the potential 
\begin{equation}
\label{V2}
  \tilde{V} = \left(\frac{\Lambda^3}{|\phi|}\right)^\kappa \VVY = C
  \Lambda^{3\kappa} |\phi|^{4/3-\kappa} \left[ \ln^2 \frac{|\phi|}{\Lambda^3} 
  + \left( \alpha -\frac{\tvac+2\pi n}{N} \right)^2 \right]~,
\end{equation}
where $C$ is a dimensionless constant. The potential $\tilde{V}$
possesses the same physical minimum at $|\phi|=\Lambda^3$ as $\VVY$,
whereas the existence and the properties of the unphysical minimum at
$\phi=0$ depend on $\kappa$. In fact, $\phi=0$ is a minimum of
$\tilde{V}$, if $\kappa<4/3$, but for $1/3<\kappa<4/3$ the first
derivative at this minimum is singular. We shall determine in the
following that the probe potential $V$ coincides with $\tilde{V}$
around the vacuum for $C=L/\Lambda$ and $\kappa=1$. 

The brane potential $V$ is given by eqn.~\eqref{V},
\begin{equation}
\label{V_new}
  V= \frac{4\pi\tau_5}{L^2} \left[ 
  \e{2\Phi} \rho \coth(2\rho+c) + \Psi(\rho) - \frac14 \e{2\Phi}(
  2a -\da) \cos(\psi-\psi_0) \right]~,
\end{equation}
where the function $\Psi$ satisfies eqn.~\eqref{dotPsi}
\begin{equation}
\label{dotPsi_new}
  \dot{\Psi}(\rho) =  2 \e{2\Phi}
    \left[\e{2h} +\frac1{16} \e{-2h} (1-a^4) \right]~.
\end{equation}
In order to discuss $V$, we must again distinguish
between the singular solutions, $c>0$, and the regular one, $c=0$,
which are qualitatively very different. Let us
start with the singular cases. We were not able to integrate eqn.\
\eqref{dotPsi_new} except for the case $c=\infty$, where
$\Psi=\e{2\Phi}(\rho-1/2)+\Psi_0$. The integration constant $\Psi_0$
plays the role of a zero-point energy, and we shall set it to zero
ensuring $V=0$ for $\rho=\hr=1/4$. Hence, we obtain
\begin{equation}
\label{Vinf}
  V_{c=\infty} = 
  4L\Lambda^3 \e{2\rho-1} \sqrt{\rho-\frac14}~,
\end{equation}
where we have used eqn.~\eqref{Phi0_interpret} to eliminate $\Phi_0$
and eqn.~\eqref{f_def} for $f(c)$. Eqn.~\eqref{Vinf} captures not only
the large $\rho$ behaviour of $V$ in all cases, but its behaviour at
the singularity is typical for the cases with $c>0$. In fact,
making the ansatz $\Psi= \e{2\Phi}[\rho-1/2+f(\rho)]$, we find from
eqn.~\eqref{dotPsi_new} the following differential equation for $f$,
\begin{equation}
\label{f_eq}
  \frac{d}{d\rho} \left(\frac{\rho f}{a \e{h}}\right) =
  -\frac{\rho}{a\e{3h}} \left(\e{2h} +\frac{a^2-1}4 \right) 
  \left[ 1-\frac1\rho \left( \e{2h} -\frac{a^2-1}4 \right) \right]~.
\end{equation}
It is not difficult to show that $f$ behaves as 
\begin{equation}
\label{fsim}
  f = f_0 \sqrt{\rho-\hr} + \frac{1-4\hr}{\hr} (\rho-\hr) +\cdots
\end{equation}
close to the singularity. Setting the integration constant $f_0$ to
zero we ensure again $V=0$ for $\rho=\hr$. Moreover, using $\e{2\Phi}\sim
(\rho-\hr)^{-1/2}$ and $2a-\da\sim\rho-\hr$, we obtain $V\sim
\sqrt{\rho-\hr}$ with a positive proportionality constant close to the
singularity. Hence, the potential has its absolute minimum at
$\rho=\hr$, but the brane probe feels an infinite attractive force at
the singularity. Furthermore, the last term in eqn.~\eqref{V_new} 
containing $\cos(\psi-\psi_0)$ exactly vanishes at the singularity, which
means that the state of lowest potential energy does not depend on the
choice of $\psi$. Both features of the minimum---invariance under variations of
$\psi$ and a singular first derivative---are in common with the
unphysical minimum of the potential $\tilde{V}$, although
the comparison does not stand up to a more quantitative analysis.

We shall discuss now the regular solution and find a
quantitative agreement between $V$ and $\tilde{V}$ close to the
vacuum for $\kappa=1$ and $C=L/\Lambda$.  
The first feature, which is different from the singular cases, is that
the coefficient in $V$ of $\cos(\psi-\psi_0)$ is strictly
negative. Hence, the vacuum must be found at $\psi=\psi_0$. (Any
multiples of $2\pi$ are irrelevant.) From eqn.~\eqref{tYM} we now
have that $\psi_0= (\tvac + 2\pi n)/N$, so that we can re-write
$\cos(\psi-\psi_0)$ as $\cos[\psi-(\tvac+2\pi n)/N]$, where $\tvac$
is the field theory vacuum angle. 
Expanding the cosine in $V$ to quadratic order and comparing it with
$\tilde{V}$, we find that $\psi=\alpha$, and we can make the following
identifications, which we expect to hold in the vicinity to the
physical minimum, 
\begin{align}
\label{phi_ident}
  C \Lambda^{3\kappa} |\phi|^{4/3-\kappa} &\approx 
  \frac{4\pi\tau_5}{L^2} \e{2\Phi} \frac18 (2a-\da)~,\\
\label{pot_ident}
  C \Lambda^{3\kappa} |\phi|^{4/3-\kappa} \ln^2
  \frac{|\phi|}{\Lambda^3} &\approx 
  \frac{4\pi\tau_5}{L^2} 
  \left[ \e{2\Phi}\left( \rho \coth(2\rho) -\frac12 a +\frac14 \da
  \right) + \Psi \right]~.
\end{align}
It is straightforward to show that the right hand side of eqn.\
\eqref{pot_ident} has exactly one local minimum, 
which is at $\rho=0$. Thus, the potential $V$ has
its minimum at $\rho=0$ and $\psi=\psi_0$, which is not chirally invariant. 

Expanding eqn.~\eqref{phi_ident} about $\rho=0$ and substituting
eqn.~\eqref{Phi0_interpret} for $\Phi_0$, we find
\begin{equation}
\label{phi_exp}
  C \Lambda^{3\kappa} |\phi|^{4/3-\kappa}  = L \Lambda^3 \left( 1
  +\frac23 \rho + \frac29 \rho^2 +\mathcal{O}(\rho^3) \right)~.
\end{equation}
Since $|\phi|=\Lambda^3$ at the minimum, we obtain $C=L/\Lambda$. 
Then, from eqn.~\eqref{phi_exp} follows
\[ \ln \frac{|\phi|}{\Lambda^3} = \frac3{4-3\kappa} \left( \frac23 \rho
+ \mathcal{O}(\rho^3) \right)~, \]
which we use to expand the left hand side of eqn.~\eqref{pot_ident},
thus obtaining 
\begin{equation}
\label{pot_exp}
  \frac{4\rho^2 L \Lambda^3}{(4-3\kappa)^2} \left( 1 +\frac23 \rho
  +\cdots\right) =
  \frac{4\pi\tau_5}{L^2} \e{2\Phi_0} 
  \left( \rho^2 + \frac23 \rho^3 +\cdots \right)~.
\end{equation}
After substituting the constant $\Phi_0$
we find agreement of the second and the third derivatives of
the potentials $V$ and $\tilde{V}$ at the minimum, if $\kappa=1$.
(The fourth derivatives do not agree). Naively one might have expected
to find $\kappa=0$, but one should bear in mind that, while $\VVY$ is
constructed from a holomorphic superfield, the probe potential is not
intrinsically holomorphic. 

Finally, we would like to point out that the result $\kappa=1$ is
supported also by a consideration of the kinetic term.
In fact, it is not difficult to show that, if we allow
for small fluctuations $\rho(x)$ and $\psi(x)$ of the brane positions,
the probe brane action gives rise also to the kinetic term 
\begin{equation}
\label{Skin}
  S_{\mathrm{kin}} = - \frac{2\pi\tau_5}{L^4} \int d^4x\, \e{2\Phi} 
  \left[\rho \coth(2\rho+c) -\frac12 a(\rho) \cos(\psi-\psi_0) \right] 
  \left[ (\partial \rho)^2 +\frac14 (\partial \psi)^2 \right]~.
\end{equation}
We consider the regular case, $c=0$, and evaluate the kinetic term in
the vicinity of the vacuum. Up to quadratic order in $\rho$ and 
$\psi-\psi_0$, we find 
\begin{equation}
\label{Skin_0}
  S_{\mathrm{kin}} \approx -\frac{2\Lambda^3}{L} \int d^4
   x\, \left[ \rho^2 +\frac14(\psi-\psi_0)^2 \right]  
  \left[ (\partial \rho)^2 +\frac14 (\partial \psi)^2 \right]~.
\end{equation}
Using $\kappa=1$ and $C=L/\Lambda$, we find to leading order from eqn.~\eqref{phi_exp}
that $\rho\approx (|\phi|/\Lambda^3 -1)/2$, so that eqn.~\eqref{Skin_0}
can be re-written (again exact to quadratic order) as 
\begin{equation}
\label{Skin_1}
  S_{\mathrm{kin}} \approx -\frac1{8L\Lambda^9} \int d^4
  x\, \left( \phi -\Lambda^3 \e{i\psi_0}\right)
  \left( \phi^\ast -\Lambda^3 \e{-i\psi_0} \right) 
  \partial_\mu \phi \partial^\mu\phi^\ast ~,   
\end{equation}
where $\phi=|\phi|\e{i\psi}$. 
In order to obtain this complexified form the value $\kappa=1$ is
crucial.

It would be very interesting to recast eqn.~\eqref{Skin_1} in
terms of a K\"ahler potential.

\acknowledgments
I am indebted to A.~Lerda for his collaboration on this
topic and  careful reading of the manuscript. 
Moreover, thanks go to M.~Bertolini, P.~Di~Vecchia,
A.~Liccardo, R.~Marotta, P.~Merlatti, F.~Nicodemi, C.~Nu\~nez and
R.~Russo, as well as the other participants of the ``gauge/gravity''
workshop during the 2003 RTN Torino winter school, for useful
discussions. Financial support by the European Commission under the
contract RTN-2000-00131 and by a fellowship from INFN is gratefully
acknowledged.

\providecommand{\href}[2]{#2}\begingroup\raggedright\endgroup

\end{document}